\documentclass[proof]{WileyASNA-v1}

\articletype{Article Type}%

\received{}
\revised{}
\accepted{}

\begin{document}

\title{Repeating FRBs reveal the secret of pulsar magnetospheric activity}

\author[aff1,aff2]{Renxin Xu*}

\address[aff1]{School of Physics and State Key Laboratory of Nuclear Physics and Technology, Peking University, Beijing 100871, China}
\address[aff2]{Kavli Institute for Astronomy and Astrophysics, Peking University, Beijing 100871, China}

\author[aff3]{Weiyang Wang}

\address[aff3]{School of Astronomy and Space Science, University of Chinese Academy of Sciences, Beijing 100049, China}

\corres{*Renxin Xu, School of Physics, Peking University, Beijing 100871, China. \\
\email{r.x.xu@pku.edu.cn\\
~
\\
DOI: 10.1002/asna.20230153}}


\abstract{
The puzzling mechanism of coherent radio emission remains unknown, but fortunately, repeating fast radio bursts (FRBs) provide a precious opportunity, with extremely bright subpulses created in a clear and vacuum-like pulsar magnetosphere. FRBs are millisecond-duration signals that are highly dispersed at distant galaxies but with uncertain physical origin(s). Coherent curvature radiation by bunches has already been proposed for repeating FRBs. The charged particles are created during central star's quakes, which can form bunches streaming out along curved magnetic field lines, so as to trigger FRBs. The nature of narrow-band radiation with time-frequency drifting can be a natural consequence that bunches could be observed at different times with different curvatures. Additionally, high linear-polarization can be seen if the line of sight is confined to the beam angle, whereas the emission could be highly circular-polarized if off-beam. It is also discussed that pulsar surface may be full of small hills (i.e., zits) which would help producing bulk of energetic bunches for repeating FRBs as well as for rotation-powered pulsars.
}



\maketitle


\section{Introduction}

It has been over half a century since the discovery of
pulsars~\citep[e.g.,][]{2017JPhCS.932a2001M}, but the underlying mechanism
responsible for pulsar coherent radio emission is
still a matter of debate. This radiative mechanism depends
on the production, acceleration, and radiation of electron-positron
$e^\pm$ pair plasma in pulsar magnetosphere, and
certainly we have known a lot about what happens: the
kinematic spin-energy of a pulsar is lost dominantly by the
current flow in its nearby magnetosphere~\citep{1969ApJ...157..869G,2006ApJ...643.1139C}; $e^\pm$-pairs
are produced and accelerated in so-called gaps, in the
mainstream, either inner-vacuum gap just above pulsar
surface \citep{1975ApJ...196...51R} or slot gap with
space charge-limited flow \citep{1979ApJ...231..854A}
and even outer-vacuum gap \citep{1986ApJ...300..522C}, to
be relevant to the binding energy of charged particles
on the surface. Nonetheless, in a view point of global
current flow, the conventional open-field line region
could be divided into annular and core sectors \citep{2004ApJ...606L..49Q}, both of which may contribute pair plasma
and thus coherent radiation. In order to understand
the non-stationary $e^\pm$-production plasma that flows out
non-homogeneously \citep{2002nsps.conf..240U}, numerical methods
with high resolution are applied to solve the mysterious
pulsar radiation problem~\citep{2006MNRAS.368.1055T}, with
extending the simulations to 2D \citep{2020PhRvL.124x5101P}.
Despite these successes, it is worth noting that the magnetospheric
activity should be subjected to the nature of pulsar surface, which is related to the big question of pulsar inner structure (i.e., the equation of state
of supra-nuclear matter at low temperature), a more
challenging problem in today’s physics and astrophysics!

It has a long history to think philosophically bulk
strong matter~\citep{2023AdPhX...837433L}, i.e., ``{\it gigantic nucleus}'' in
Landau’s words presented just in the late age of developing
quantum mechanics~\citep{Landau1932}.
Now it is well known that pulsar-like compact stars are such objects in
reality, but unfortunately/fortunately, the nature of pulsars
has remained still a mystery even after more than 90
years \citep{Xu2003}, supposed to be the first big one to be
solved in the era of gravitational-wave astronomy. The pulsar
matter could be extremely isospin-unsymmetrical so
that normal baryonic crust would be necessary to cover
(i.e., normal neutron star model), but could also probably
consist of 2- or 3-flavored itinerant quarks (i.e., quark star
model). In view of charge and flavor symmetries of light
quarks (up, down, and strange) and no-perturbativity of
the fundamental strong interaction at pressure free and at
zero temperature, it is proposed that the building blocks of
pulsar matter could be “strangeons”, an analogy of atomic
nucleons but with negative strangeness \citep{Xu2003,LX2017}. This strangeon star model could be successful in
explaining, such as massive pulsars and low tidal deformability \citep{2019EPJA...55...60L}, the spin-irregularity of glitches (e.g., \citealt{2023MNRAS.tmp.1618L}), and huge free energy for bursts and
flares \citep{2023arXiv230519687C,2006MNRAS.373L..85X}, and we are concerning
here about its implications to pulsar magnetospheric
activity.

Two immediate consequences of strangeon matter on
pulsar surface should be of ``miser'' and ``zits''.
(1) {\it Miser}.
A bare strangeon star looks like “a miser” because both
positively and negatively charged particles can hardly be
extracted from the surface so that an inner-vacuum gap
works naturally there \citep{1999ApJ...522L.109X,2011MNRAS.414..489Y}.
(2) {\it Zits}.
Although the interactions between nucleons and
strangeons could both be ``Van der Waals''-like, a straneon
is more akin to a classical particle than quantum
wave due to its high mass, $m_{\rm st}$, as indicated by its Compton
wavelength of $\lambda\simeq \hbar/(m_{\rm st}c)$. It was then conjectured
that condensed strangeon-matter could be in a classical
solid state \citep{Xu2003}, and a bare strangeon star surface
may have small hills (i.e., ``zits'') where the induced electric
fields parallel to magnetic field-lines should be remarkably
stronger than that anywhere else. Nonetheless, how can
we find any clues to the strangeon surface? The booming
research on fast radio bursts (FRBs) could be one of the
examples.

In short, FRBs are bright radio transients prevailing in
the universe with milli-second duration. The field has witnessed
a rapid increase in the frontiers of observations and theories since the first FRB discovery \citep{2007Sci...318..777L}.
Without doubt, the key issue in the field is to understand
FRB's radiative mechanism and further the underlying
physics relevant to FRB's central engine.

FRB sources can fall into two groups: repeaters and
apparently non-repeating ones. However, whether all
FRBs can repeat is still an open question even the two
groups show some noticeable differences in the burst morphology
\citep{2021ApJ...923....1P}. The sources are thought to
be cosmological origin due to their dispersion measures
(DMs) in excess of the Galactic values, and it was confirmed
after the first repeater was localized in its host
galaxy \citep{2017ApJ...843L...8B,2017Natur.541...58C,2017ApJ...834L...8M}. Besides, an FRB-like burst was found in a
Galactic compact star (soft gamma ray repeater, SGR),
so-called ``magnetar'' \citep{2020Natur.587...59B,2020Natur.587...54C}, which is named as SGR
J1935+2154, and the burst is temporally associated with
an X-ray burst \citep{2021NatAs...5..378L,2021NatAs...5..372R,2021NatAs...5..401T}. This discovery suggests that at least some FRBs
could be created by extragalactic magnetars. That is not
surprising as pulsed radio emission, thought much weaker,
are sometimes detected from these objects after out-bursts,
for example, the fifth radio-loud Galactic magnetar Swift
J1818.0-1607 ~\citep{2020ApJ...896L..30E}.

Some intriguing properties have been found in repeating
FRBs which are surely meaningful to study their origin.
For instance, the ``sad trombone'' spectral structure,
that is, subpulses with higher frequency arrive earlier than
those with lower frequencies, has been found in some
repeating FRBs (e.g., \citep{2019ApJ...876L..23H}). Careful polarization measurements
can give more physical information
about their magnetospheric origin. In general, most bursts
are dominated by linear polarization (LP) and present a
flat polarization position angle (PA) across each pulse, but
there are also some bursts that have circular polarization
(CP) fractions larger than 50\% and even up to 75\% \citep{2022Natur.609..685X}. FRB 180301, a repeater, exhibits varying PAs
across the burst envelope \citep{2020Natur.586..693L}, which is reminiscent
of an S-shaped PA across the burst envelope of a
pulsar. All these might reveal the underlying mechanism
of coherent radio emission in pulsar magnetosphere, and
a strangeon matter surface could even be needed.
 
In this paper, we review the trigger and radiation
mechanism of repeating FRBs. In Section 2, we focus
on starquake as the trigger mechanism. In Section 3, we
demonstrate that the drifting pattern and the variety of
radio polarization (both linear and circular) can be well
understood within the frame work of coherent curvature
radiation by bunches. In Section 4, high-tension point discharge
at ``zits'' on a strangeon surface is discussed. We
summarize our results in Section 5.

\section{Quake-induced repeating FRBs?}

Quakes could occur in pulsar-like compact objects, either
for normal neutron stars (magnetras) with ultra-strong
magnetic fields when the internal magnetic field exceeds
a threshold stress, or for strangeon stars when slow elastic
loading develops to a critical point so as to release unstably.
An FRB could be triggered by a quake, though no final conclusion
is obtained about the free energy:magnetic energy
for magnetars or elastic/gravitational energy for strangeon
stars. Both processes may let a lot of energy release from
the central star into the magnetosphere so that charged
particles are then accelerated and stream out along the
magnetic field lines.

From the statistical side, nonlinear dissipative systems
can show self-organized criticality (SOC) behaviors \citep{2011soca.book.....A}. The star whether a pulsar-like compact
star or the Earth, can build up stresses that make
the crust crack and adjust the stellar shape in order to
reduce its deformation. It is then expected that pulsar-like
compact star activities (e.g., magnetar quake) may exhibit
characteristics of self-organized criticality, as has been
observed in earthquakes.
Quakes from pulsar-like compact
stars can share similar properties with earthquakes.

In 2017, 14 bursts above the threshold of 10 sigma
were detected by Breakthrough Listen Digital Backend
with the C-band receiver at the Green Bank Telescope \citep{2017ATel10675....1G}. We found that the burst sequence
exhibits some earthquake-like behaviors such as the
power-law burst energy distribution, which is reminiscent
of the Gutenberg-Richter law, a typical Earthquake
behavior \citep{2017JCAP...03..023W,2018ApJ...852..140W,2019ApJ...882..108W}. The temporal behavior of the burst sequence
satisfies the Omori law, which interprets the time decay
of the seismicity rates of an aftershock sequence. The
timescale for this process is $\sim L/v_A\approx1-10$ ms, where
$L$ is the scale of the reconnection-unstable zone and $v_A$
is the Alfv{\'e}n velocity. The $e^{\pm}$-pair production is excited
during the starquake, and electrons or positions (as a
leading charge in the following discussion) are suddenly
accelerated to ultra-relativistic velocity, creating coherent
radio emissions. The high burst rates by plate collisions
in central star’s crusts could be consistent with the observation
of FRB 20121102A, with maximum value reaching $122\,\rm h^{-1}$~\citep{2022MNRAS.517.4612L}.

\section{Coherent curvature radiation by magnetospheric bunches}

For FRBs, an extremely coherent radiation mechanism is
required to explain the high brightness temperature.
The trajectories of charged particles are tracked by magnetic field lines because the vertical momentum perpendicular
to the field line damps very fast. Curvature radiation can
be created by the streaming charged bunches.

Charged bunches can be formed if their sizes are
smaller than the half wavelength of the emission so
that the luminosity is proportional to $N^2_e$ rather than $N_e$,
where $N_e$ is the number of charges inside the bunch. The
flux of curvature radiation is peaked at $\nu_c = 3c\gamma^3/(4\pi\rho)=0.7\gamma^3_2\rho_7^{-1}$ GHz, where $\gamma$ is the Lorentz factor of bunches,
$c$ is the speed of light, and $\rho$ is the curvature radius. In
order to match the observed FRB frequency, the emission
region should be from several tens to several hundred stellar
radii with a required Lorentz factor in order of several
hundreds.
An FRB can be seen when the line of sight (LOS)
sweeps across the field lines where the bunches appear
coincidentally.

If one can observe two or more separated bulk of
bunches, there would be two or more sub-pulses. We
assume that the emitting two bulks are generated at the
same time so that the observed time interval of subpulses
is mainly caused by the geometric delay. By considering
an axisymmetric magnetic configuration, one can obtain
the geometric time delay for emissions from two bulk of
bunches:
\begin{equation}
\Delta t_{\rm geo}\simeq\frac{\Delta r}{c}\left[\left(1+\frac{1}{2\gamma^2}\right)I_n(\theta)-\cos\theta_p\right],
\label{eq:tgeo2}  
\end{equation}
where $\Delta r$ denotes the distance between the points where
emission can sweep the LOS at two magnetic field lines and $n$ denotes magnetic configuration ($n=1$ for dipole) \citep{2022ApJ...927..105W}.
Here the subscript "1" denotes the bulk at lower height and "2" for at higher height.
The result of equation (1) is always positive independent with magnetic configuration of $n$.
Therefore, the pulses seen later travel farther into the less-curved part of
the magnetic field lines, thus emitting at lower frequencies,
matching the observed downward drifting pattern \citep{2019ApJL..876L..15W}.

The observed amplitude can be demonstrated by the summation of the curvature radiation amplitude of individual particles since FRB emissions is coherent.
Basically, $A_{\|}$ and $A_{\perp}$ are defined as two orthogonal polarized components of the amplitude along $\boldsymbol{\epsilon}_{\|}$ and $\boldsymbol{\epsilon}_{\perp}$, where $\boldsymbol{\epsilon}_{\|}$ is the unit vector pointing to the center of the instantaneous circle, and $\boldsymbol{\epsilon}_{\perp}=\boldsymbol{n} \times \boldsymbol{\epsilon}_{\|}$ is defined \citep{1998clel.book.....J}.
The two amplitudes of bunch are the summation of individual particles which are given by
\begin{equation}
\begin{aligned}
A_{\|} & \simeq \frac{i 2}{\sqrt{3}} \frac{\rho}{c}\left(\frac{1}{\gamma^{2}}+\varphi^{2}+\chi^{2}\right) K_{\frac{2}{3}}(\xi)\\
&+\frac{2}{\sqrt{3}} \frac{\rho}{c} \chi\left(\frac{1}{\gamma^{2}}+\varphi^{2}+\chi^{2}\right)^{1 / 2} K_{\frac{1}{3}}(\xi), \\
A_{\perp} & \simeq \frac{2}{\sqrt{3}} \frac{\rho}{c} \varphi\left(\frac{1}{\gamma^{2}}+\varphi^{2}+\chi^{2}\right)^{1 / 2} K_{\frac{1}{3}}(\xi),
\end{aligned}
\end{equation}
where $\chi$ is the angle between the considered trajectory and trajectory at $t = 0$, $\varphi$ is the angle between the LOS and trajectory plane, and $K_\nu$ is the modified Bessel function.

\begin{figure*}[h] 
\begin{minipage}{0.32\linewidth}
\vspace{3pt}
  \centerline{\includegraphics[width=\textwidth]{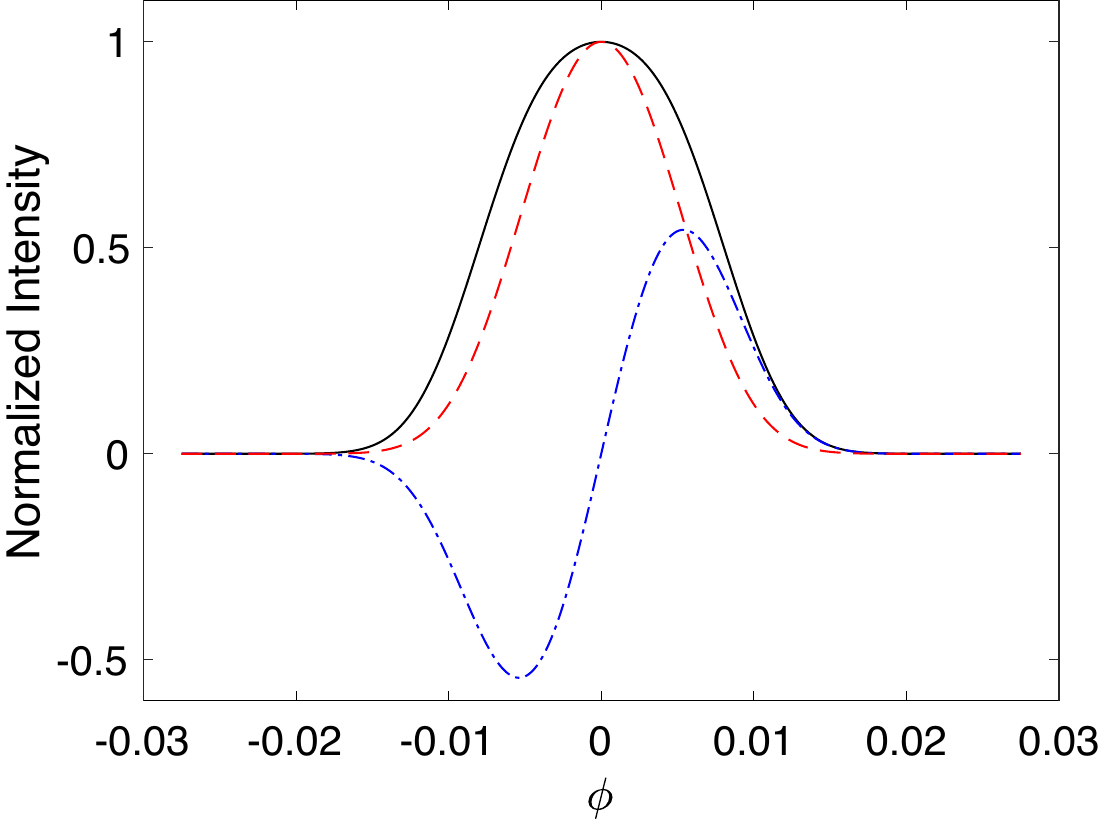}}
\centerline{(a)}
\end{minipage}
\begin{minipage}{0.32\linewidth}
\vspace{3pt}
\centerline{\includegraphics[width=\textwidth]{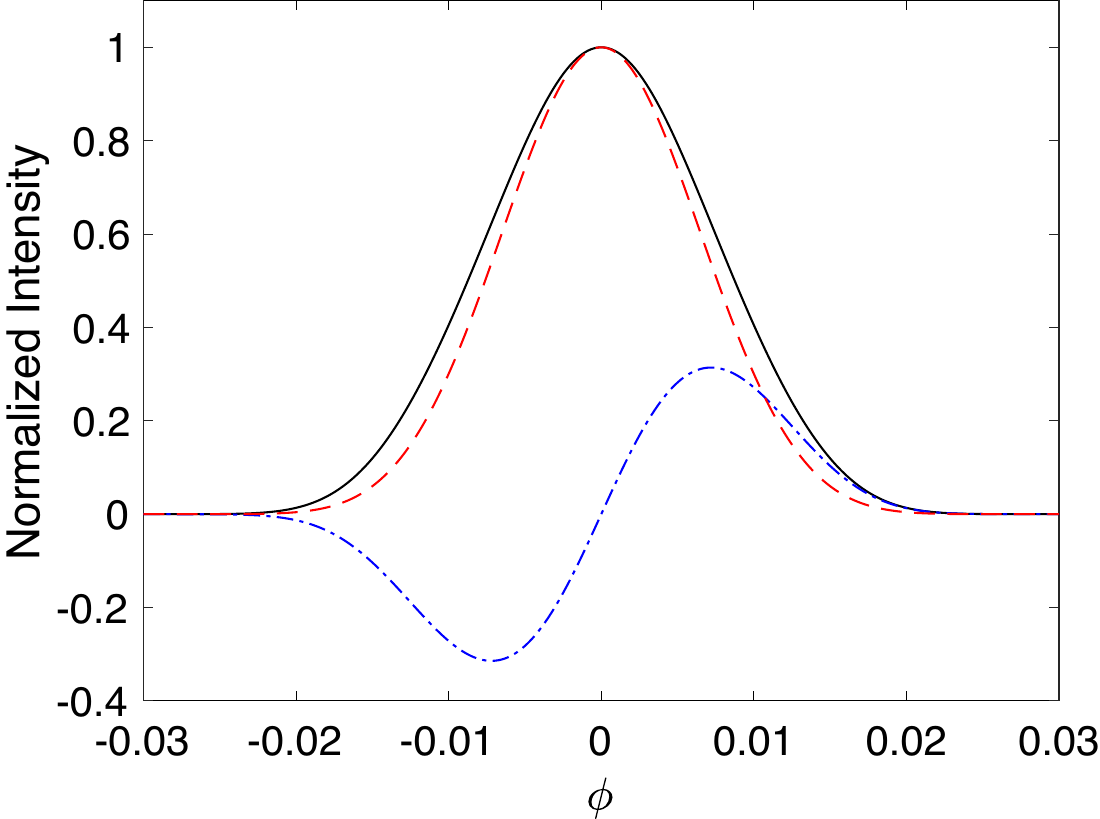}}
 \centerline{(b)}
\end{minipage}
\begin{minipage}{0.32\linewidth}
\vspace{3pt}
\centerline{\includegraphics[width=\textwidth]{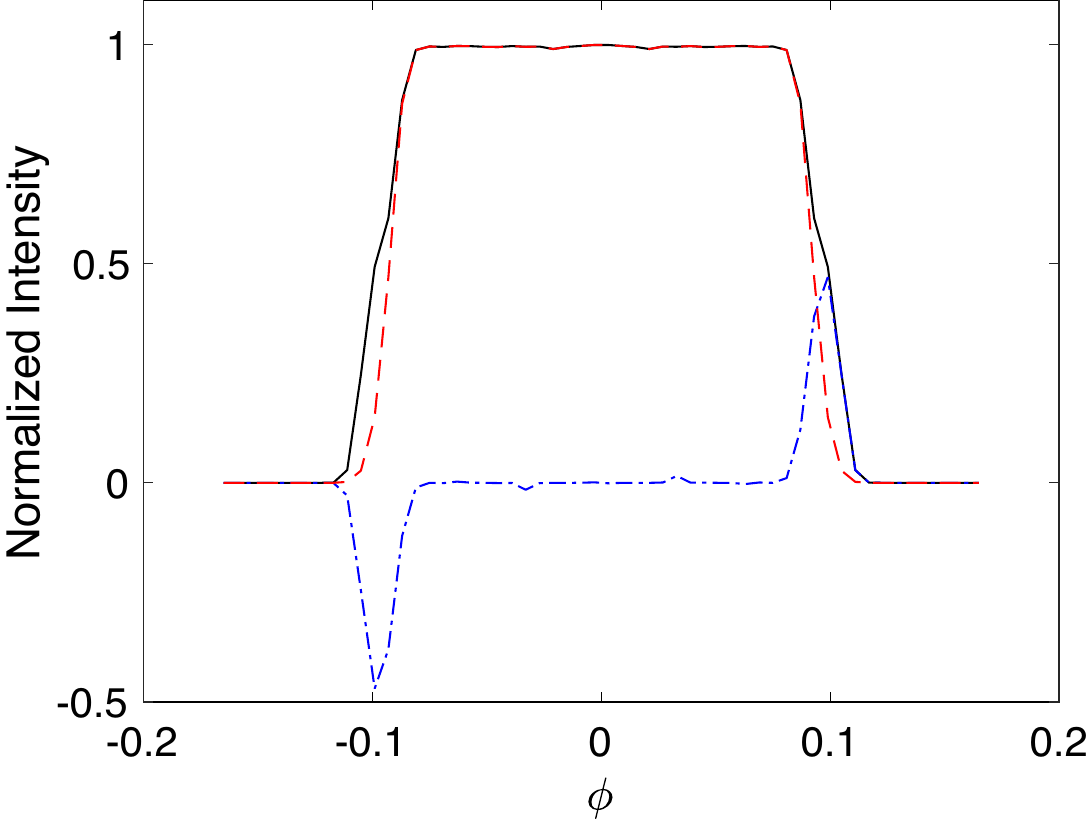}}
\centerline{(c)}
\end{minipage}
\caption{Simulated polarization profile and PA across the burst envelope for different frequencies: (a) $\varphi_t=0.1/\gamma$; (b) $\varphi_t=1/\gamma$; (c) $\varphi_t=10/\gamma$. The Normalized intensities are plotted in black solid lines. The LP fractions are plotted in red dashed lines. The CP fractions are plotted in blue dotted-dashed lines.}
\label{fig4}
\end{figure*}

According to Equation (2), we consider three cases with different bunch scale ($\varphi_t=0.1/\gamma$, $\varphi_t=1/\gamma$, $\varphi_t=10/\gamma$, where $\varphi_t$ is the half opening angle of bunch).
The polarization profiles across burst envelope are plotted in Figure \ref{fig4}.
Let us define that the LOS is confined to the radiation beam as on-beam, whereas off-beam.
The beam angle for a bunch is defined as $\theta_b = \varphi_t + \theta_c$ , where $\theta_c$ is spread angle of emission for single charge \citep{2022MNRAS.517.5080W}.
Emission from an ultrarelativistic particle is mainly confined in a conal region (spread angle).
The spread angle for curvature radiation of a single charge is $1/\gamma$ when $\nu=\nu_c$.

As shown in Figure \ref{fig4}, emissions for all three cases have highly LP fractions if the LOS is inside the beam (on-beam), and the CP fraction becomes significant when the LOS is off-beam.
Waves are of left-CP at $\phi<0$ but change to right-CP after the LOS sweeps across the central axis of the bunch.
For on-beam case, there is a large phase space where the summation of $A_\perp$ cancels out, so that the emission has roughly $100\%$-LP fraction.
In a word, the emission can only become highly circular polarized for the off-beam cases.

Additionally, we investigate two propagation scenarios,
including Faraday conversion and absorption in closed
field line region. Faraday conversion appears at a magnetized
plasma medium which has B component perpendicular
to the LOS. Conversion between LP and CP can be seen due to the difference in group velocity between
O mode and X mode photons. However, the LP and CP
fractions oscillate with $\lambda^3$ inconsistent with the observation
of $\lambda^2$-oscillation in FRB 20201124A. Outcome emissions
with a high CP still need the income waves to have
large CP fractions \citep{2022SCPMA..6589511W}. Another scenario is
the absorption effect. Right CP waves are likely to be optically
thick for a rapidly rotating neutron star with a strong
magnetic field \citep{2021RAA....21...29L}. Emissions with highly LP
fractions are required to be emitted from at least higher
than the absorption region, leading to a time delay between
LP and CP bursts

\section{``Zits'' on pulsar's surface?}

\begin{figure}[h!]
\centering
\includegraphics[scale=0.3]{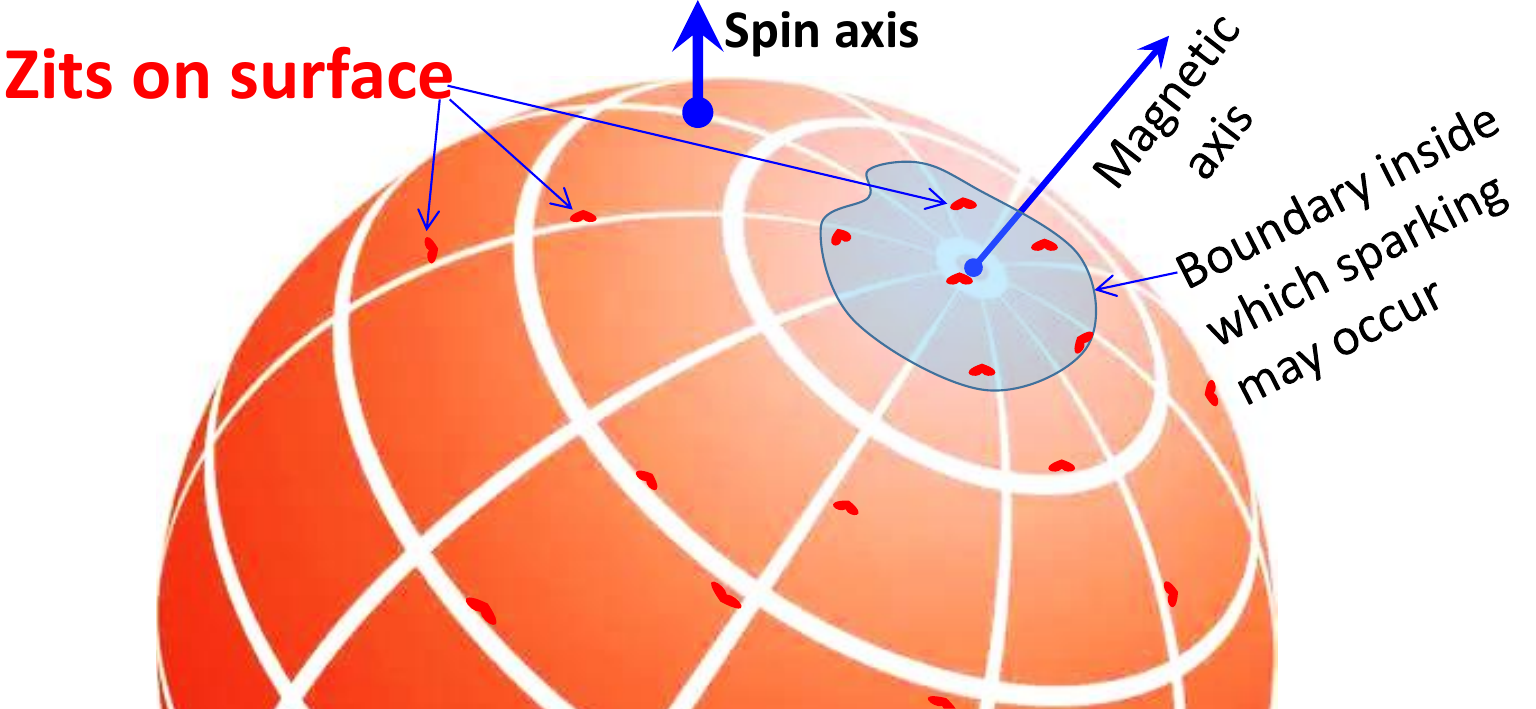}
\caption{A bare strangeon star should not have a smooth surface, but be covered with pimples, i.e. the zits illustrated here. These small hills on pulsar surface might be responsible for the magnetospheric activity relevant to coherent radio emission.
} %
\label{f2}
\end{figure}

It is multi-motivated to study mountain building of solid
strangeon stars \citep{2011arXiv1104.5287Y}, and the maximum
height, hmax, ofmountains on strangeon star surface could
be estimated by an order-of-magnitude calculation with $\mu\sim \rho g h_{\rm max}$,
\begin{equation}
h_{\rm max} \sim (7\times 10^2~{\rm cm})~\mu_{32}\rho_{15}^{-1}M_1^{-1}R_6^2,
\end{equation}
where $\mu=(10^{32}~{\rm ergs~cm^{-3}})~\mu_{32}$ is the shear modulus, $\rho=(10^{15}~{\rm g~cm^{-3}})~\rho_{15}$ is the surface density, and the mass and radius are, $M=M_\odot~M_1$ and $R=(10^6~{\rm cm})~R_6$, respectively.
Soon after core-collapse supernova, a hot strangeon star
should be in a liquid state at temperate of a few 10
MeV \citep{2011SCPMA..54.1541D,2009ScChG..52..315X,2017RAA....17...92Y},
but it would be solidified at temperate of a few 0.1 MeV less
than hours. Initially, the surface should be slippery after
quenching of liquid-solid phase transition, and the landscape
may comprehend rolling country and plains, with
differences of vertical altitude $<h_{\rm max}$.
However, the surface of an aged strangeon star could be full of small hills (i.e., ``zits''-like, depicted in Fig.~\ref{f2}), especially in the polar cap region, due to the bombardment of TeV-$e^+$/$e^+$ on the polar cap, with the kinematic energy in the center of mass of $\gtrsim$ GeV~\citep{2000ASPC..202..479X}.
Frankly speaking, strangeon star-quake to be responsible for pulsar glitch
observed~\citep[e.g.,][]{2023MNRAS.tmp.1618L}~would additionally be a strong pressed orogenic movement to create zits.
It worth noting that the strong zits~\footnote{%
The typical height of zits could be $\sim (10^{-3}-10^{-1})h_{\rm max}$, i.e., several millimeters to decimeters, by the experience of comparing mountain peaks and the Himalayas on the Earth.
} on strangeon matter surface would keep a long time, while the electromagnetic zits on naked neutron star~\citep[e.g.,][]{2003A&A...407..315G,2004ApJ...603..265T,2023arXiv230602537M} surface could hardly stand up against the collisions of TeV-$e^\pm$ pairs.

It is well known that $e^\pm$-pairs are accelerated and produced
by the spin-induced electric fields, $E_\|$, which could
be too weak to be active for slow and low-magnetic field
pulsar, but $E_\|$-field could be enhanced by ($10-10^3$) times
at the peak of hill, so-called point-discharging. This may
imply that the inner-vacuum gap sparking occurs regularly
in “rolling country and plain” regions (manifested
as ordered sub-pulse drifting), but has priority to happen
irregularly around zits. Alternatively, regular and irregular
sub-pulse drifting might be the result of quasi-periodic
discharging around several zits. In this picture, it is very
natural for $e^\pm$-pairs plasma to flow out nonstationarily
and inhomogeneously in case of the point discharges. This
could be good news for coherent radio emission of pulsar~\citep{2002nsps.conf..240U}.

Observational facts may hint at the existence of zits on
pulsar’s face. A single pulse observation with China’s FAST
could be evidence for a rough surface of PSR B2016+28 via
the detection that the modulation period along pulse series
is positively correlated to the separation between two adjacent
sub-pulses~\citep{2019SCPMA..6259505L} as higher $E_{||}$ would led to faster drifting of sub-pulses.
The unusual arc-like structure of the bright pulsar PSR B0329+54~\citep{2007MNRAS.379..932M}, that is, the distinct core-weak patterns~\citep{2023MNRAS.520.4173W}, might result from the suppression of ``core''-sparking by point discharge at zits outside the core. Recently,
the asymmetric sparking points located away from the
magnetic pole of the the whole-pulse-phase pulsar PSR
B0950+08~\citep{2022MNRAS.517.5560W}, as revealed by the polarization measurements of both integrated and single pulse profiles~~\citep{2023arXiv230807691W}, would be new evidence for zits on a strangeon's face. Certainly, further studies of radio
singe pulses, especially with the highly-sensitive FAST,
are surely welcome to find solid evidence for zits on face.

What is the magnetospheric difference between regular
pulsars and repeating FRBs if pulsar-like compact
stars are responsible for both kinds of the extremely
coherent radio emission? The essential difference may
arise from the fact that regular pulsars are basically rotation-powered, while repeating FRBs could probably be
powered by a stellar activity via either a starquake or magnetic
re-connection near the surface, both being able to
contribute mountain-building movement. An FRB object
is usually below the deathline at ordinary times, exhibiting
a vacuum-like clean magnetosphere, but $e^\pm$-plasma
can occasionally erupt from the star through sparking
around zits during an active period. This may result in a
higher coherence and thus a bright emission of FRBs.

\section{Summary}

Pulsar-like compact objects are exceptionally focused in
the era of multi-messenger astronomy, with two mysteries
to be solved: the radiative mechanism of coherent
radio emission and the equation of state of cold matter
at supra-nuclear density, both studies of which have a
very long history. It is explained that these two problems
could be internally related: the analysis of radiation features
could play an important role in revealing the surface
of central star and thus the nature of dense matter. We
propose that, because of the high rigidity of strangeon
matter, pulsar surface may be full of small hills (i.e., zits)
which would help producing bulk of energetic bunches
for repeating FRBs as well as for rotation-powered pulsars.
More observational examinations should be carried out in
order to clarify the issues of whether pulsars are neutrons
or strangeon stars.

Even though there are hundreds of FRB sources have
been discovered and dozens of them can repeat the physical
origin(s) of FRBs are still unknown. We review that
coherent curvature radiation by bunches as the radiation
mechanism for repeating FRBs. The spectra-temporal
pulse-to-pulse properties can be well understood within
the framework of curvature radiation. A downward drifting
pattern is a natural consequence that bulks observed at
an earlier time were always emitted in a more-curved part
of field line.

FRBs can exhibit a wide variety of polarization properties,
not only between sources but also from burst to burst
for a same one. Within the coherent curvature radiation,
high LP would appear when the line of sight is inside the
emission beam (the on-beam case), whereas no-zero high
CP would be presented when it is outside (the off-beam
case). By considering the bulk shape and pulsar’s spin,
one can only observe part of beam so that a wide variety
of polarization can be detected \citep{2022MNRAS.517.5080W}.
Cyclotron resonance as a propagation effect can result in
the absorption of one CP mode photons at a low altitude
region of the magnetosphere, and a circularized FRB
should thus be emitted from a high-altitude region. Faraday
conversion can produce CP with Stokes parameters
oscillation, but the mean CP depends only on the income
wave.

\section*{Acknowledgments}
The authors would like to thank those involved in the continuous discussions in the pulsar group at Peking University. This work is supported by the National SKA Program of China (No. 2020SKA0120100) and the Strategic Priority Research Program of the Chinese Academy of Sciences (No. XDB0550300).

\bibliography{ref}%

\end{document}